\def\ie{{\it i.e.}}

\def\~{{$\tilde{\phantom{a}}$}}

\documentclass [12pt] {article}
\usepackage{epsfig}
\usepackage{color}

\def\thebibliography#1{\section{References}\markboth
 {REFERENCES}{REFERENCES}\list
 {[\arabic{enumi}]}{\settowidth\labelwidth{[#1]}\leftmargin\labelwidth
 \advance\leftmargin\labelsep
 \usecounter{enumi}}
 \def\newblock{\hskip .11em plus .33em minus -.07em}
 \sloppy
 \sfcode`\.=1000\relax}
\def\upcite#1{\raise6pt\hbox{\scriptsize
\cite{#1}}}
  \def\lsim{\mathrel {\vcenter {\baselineskip 0pt \kern 0pt
    \hbox{$<$} \kern 0pt \hbox{$\sim$} }}}
    \def\gsim{\mathrel {\vcenter {\baselineskip 0pt \kern 0pt
    \hbox{$>$} \kern 0pt \hbox{$\sim$} }}}

\def\hline{\noalign{\hrule \vskip2pt}}

\def\|{\ifmmode\Vert\else \char`\|\fi}
\ifx\oldzeta\undefined                          
  \let\oldzeta=\zeta                            
  \def\zzeta{{\raise 2pt\hbox{$\oldzeta$}}}     
  \let\zeta=\zzeta                              
\fi

\ifx\oldchi\undefined                           
  \let\oldchi=\chi                              
  \def\cchi{{\raise 2pt\hbox{$\oldchi$}}}       
  \let\chi=\cchi                                
\fi

\def\frac#1#2{{#1 \over #2}}

\def\half{\ifinner {\scriptstyle {1 \over 2}}
   \else {1 \over 2} \fi}

\def\simge{\mathrel{%
   \rlap{\raise 0.511ex \hbox{$>$}}{\lower 0.511ex \hbox{$\sim$}}}}
\def\simle{\mathrel{
   \rlap{\raise 0.511ex \hbox{$<$}}{\lower 0.511ex \hbox{$\sim$}}}}

\def\buildchar#1#2#3{{\null\!                   
   \mathop#1\limits^{#2}_{#3}                   
   \!\null}}                                    
\def\overcirc#1{\buildchar{#1}{\circ}{}}

\def\slashchar#1{\setbox0=\hbox{$#1$}           
   \dimen0=\wd0                                 
   \setbox1=\hbox{/} \dimen1=\wd1               
   \ifdim\dimen0>\dimen1                        
      \rlap{\hbox to \dimen0{\hfil/\hfil}}      
      #1                                        
   \else                                        
      \rlap{\hbox to \dimen1{\hfil$#1$\hfil}}   
      /                                         
   \fi}                                         %

\def\subrightarrow#1{
  \setbox0=\hbox{
    $\displaystyle\mathop{}
    \limits_{#1}$}
  \dimen0=\wd0
  \advance \dimen0 by .5em
  \mathrel{
    \mathop{\hbox to \dimen0{\rightarrowfill}}
       \limits_{#1}}}                           

\def\overlay#1#2{\ifmmode%
\setbox0=\hbox{$#1$}%
\setbox1=\hbox to\wd0{\hss$#2$\hss}\else%
\setbox0=\hbox{#1}%
\setbox1=\hbox to\wd0{\hss#2\hss}\fi%
#1\hskip-\wd0\box1 }

\def\pmb#1{\leavevmode\setbox0=\hbox{#1}%
\kern-.02em\copy0\kern-\wd0
\kern.04em\copy0\kern-\wd0
\kern-.02em\raise.04em\box0 }

\def\vereq#1#2{\lower3pt\vbox{\baselineskip1.5pt \lineskip1.5pt
\ialign{$\m@th#1\hfill##\hfil$\crcr#2\crcr\sim\crcr}}}

\def\tensor#1{\protect\@ontopof{#1}{\leftrightarrow}{1.15}\mathord{\box2}}
\def\overstar#1{\protect\@ontopof{#1}{\ast}{1.15}\mathord{\box2}}
\def\overdots#1{\protect\@ontopof{#1}{\cdots}{1.0}\mathord{\box2}}
\def\overcirc#1{\protect\@ontopof{#1}{\circ}{1.2}\mathord{\box2}}
\def\loarrow#1{\protect\@ontopof{#1}{\leftarrow}{1.15}\mathord{\box2}}
\def\roarrow#1{\protect\@ontopof{#1}{\rightarrow}{1.15}\mathord{\box2}}

\def\@ontopof#1#2#3{%
{\mathchoice
{\@@ontopof{#1}{#2}{#3}\displaystyle\scriptstyle}%
{\@@ontopof{#1}{#2}{#3}\textstyle\scriptstyle}%
{\@@ontopof{#1}{#2}{#3}\scriptstyle\scriptscriptstyle}%
{\@@ontopof{#1}{#2}{#3}\scriptscriptstyle\scriptscriptstyle}%
}%
}

\def\@@ontopof#1#2#3#4#5{%
\setbox0=\hbox{$#4#1$}%
\setbox1=\hbox{$#5#2$}%
\setbox2=\hbox{}\ht2=\ht0 \dp2=\dp0 %
\ifdim\wd0>\wd1 %
\setbox1=\hbox to\wd0{\hss\box1\hss}%
\mathord{\rlap{\raise#3\ht0\box1}\box0}%
\else   %
\setbox1=\hbox to.9\wd1{\hss\box1\hss}%
\setbox0=\hbox to\wd1{\hss$#4\relax#1$\hss}%
\mathord{\rlap{\copy0}\raise#3\ht0\box1}%
\fi
}%

\def\lambdabar{\protect\@lambdabar}
\def\@lambdabar{%
\relax
\bgroup
\def\@tempa{\hbox{\raise.73\ht0
\hbox to0pt{\kern.25\wd0\vrule width.5\wd0
height.1pt depth.1pt\hss}\box0}}%
\mathchoice{\setbox0\hbox{$\displaystyle\lambda$}\@tempa}%
{\setbox0\hbox{$\textstyle\lambda$}\@tempa}%
{\setbox0\hbox{$\scriptstyle\lambda$}\@tempa}%
{\setbox0\hbox{$\scriptscriptstyle\lambda$}\@tempa}%
\egroup
}

\def\corresponds{{\lower.2ex\hbox{=}}{\rm\kern-.75em^\triangle}}
\def\succsim{\succ\kern-.9em_\sim\kern.3em}
\def\precsim{\prec\kern-1em_\sim\kern.3em}
\def\slantfrac#1#2{\kern1em^{#1}\kern-.3em/\kern-.1em_{#2}}

\begin{document}

\begin{center}
{\Large\bf A mechanical model that exhibits a gravitational
critical radius}
\\

\medskip

Kirk T.~McDonald
\\
{\sl Joseph Henry Laboratories, Princeton University, Princeton, NJ 08544}
\\
(Dec.\ 2, 1998)
\end{center}

\section{Problem}

A popular model at science museums (and also a science toy \cite{vortx})
 that illustrates how curvature
can be associated with gravity consists of a surface of revolution
$r = -k/z$ with $z < 0$ about a vertical axis $z$.  
The curvature of the surface, combined with the
vertical force of Earth's gravity, leads to an inward horizontal
acceleration of $kg/r^2$ for a particle that slides freely on the
surface in a circular, horizontal orbit.

Consider the motion of a particle that slides freely on
 an arbitrary surface of revolution, $ r = r(z) \geq 0$, defined by a
continuous and differentiable function on some interval of $z$.  The surface
may have a nonzero minimum radius $R$ at which the slope $dr/dz$ is infinite.
Discuss the character of oscillations of
the particle about circular orbits to deduce a condition that there be
a critical radius $r_{\rm crit} > R$, below which the orbits are unstable.
That is, the motion of a particle with $r < r_{\rm crit}$ rapidly
leads to
excursions to the minimum radius $R$, after which the particle falls
off the surface.


Give one or more examples of analytic functions $r(z)$ that exhibit a
critical radius as defined above.
These examples provide a mechanical analogy as to how departures
of gravitational curvature from that associated with a $1/r^2$
force can lead to a characteristic radius inside which all motion
tends toward a singularity.

\section{Solution}

We work in a cylindrical coordinate system $(r,\theta,z)$ with the
$z$ axis vertical.  It suffices to consider a particle of unit mass.

In the absence of friction, there is no torque on a particle about
the $z$ axis, so the angular momentum component $J = r^2 \dot\theta$
about that axis is a constant of the motion, where $\dot{\phantom{a}}$ indicates
differentiation with respect to  time.

For motion on a surface of revolution $r = r(z)$, we have $\dot r = r' \dot z$,
where $'$ indicates differentiation with respect to $z$.
Hence, the kinetic energy can be written
\begin{equation}
T = {1 \over 2} (\dot r^2 + r^2 \dot\theta^2 + \dot z^2)
= {1 \over 2} [\dot z^2 (1 + r^{'2}) + r^2 \dot\theta^2].
\label{eq2}
\end{equation}
The potential energy is $V  = gz$.
Using Lagrange's method, the equation of motion associated with
the $z$ coordinate is
\begin{equation}
\ddot z (1 + r^{'2}) + \dot z^2 r r^{''} = -g + {J r' \over r^3}.
\label{eq3}
\end{equation}

For a circular orbit at radius $r_0$, we have
\begin{equation}
r_0^3 = {J^2 r'_0 \over g}.
\label{eq4}
\end{equation}
We write $\dot\theta_0 = \Omega$, so that $J = r_0^2 \Omega$.

For a perturbation about this orbit of the form
\begin{equation}
z = z_0 + \epsilon \sin\omega t,
\label{eq5}
\end{equation}
we have, to order $\epsilon$,
\begin{eqnarray}
r(z) & \approx & r(z_0) + r'(z_0)(z - z_0)
\nonumber \\
& = & r_0 + \epsilon r'_0 \sin\omega t,
\\
r' & \approx & r'_0 + \epsilon r^{''}_0 \sin\omega t,
\\
{1 \over r^3} & \approx & {1 \over r_0^3} \left( 1 - 3 \epsilon
\sin\omega t {r'_0 \over r_0} \right).
\label{eq10}
\end{eqnarray}

Inserting (\ref{eq5}-\ref{eq10}) into (\ref{eq3}) and keeping terms only to
order $\epsilon$, we obtain
\begin{equation}
- \epsilon \omega^2 \sin\omega t (1 + r_0^{'2}) \approx
-g + {J^2 \over r_0^3} \left( r'_0 
- 3 \epsilon \sin\omega t {r_0^{'2} \over r_0} 
+ \epsilon \sin\omega t\, r_0^{''} \right).
\label{eq11}
\end{equation}  
From the zeroeth-order terms we recover (\ref{eq4}), and from the
order-$\epsilon$ terms we find that
\begin{equation}
\omega^2 = \Omega^2 {3 r_0^{'2} - r_0 r_0^{''} \over 1 + r_0^{'2} }.
\label{eq12}
\end{equation}
The orbit is unstable when $\omega^2 < 0$, \ie, when
\begin{equation}
r_0 r_0^{''} > 3 r_0^{'2}.
\label{eq13}
\end{equation}
This condition has the interesting geometrical interpretation (noted by a
referee) that the orbit is unstable wherever
\begin{equation}
(1/r^2)'' < 0,
\label{eq13a}
\end{equation}
\ie, where the function $1/r^2$ is concave inwards.
 
For example, if $r = -k/z$, then $1/r^2 = z^2/k^2$ is concave outwards, 
$\omega^2 = J^2/(k^2 + r_0^4)$, and
there is no regime of instability.

We give three examples of surfaces of revolution that satisfy condition
(\ref{eq13a}).

First, the hyperboloid of revolution defined by
\begin{equation}
r^2 - z^2 = R^2,
\label{eq1}
\end{equation}
where $R$ is a constant.  Here, $r'_0 = z_0/r_0$,
$r_0^{''} = R^2/r_0^3$, and
\begin{equation}
\omega^2 = \Omega^2 {3z_0^2 - R^2 \over 2 z_0^2 + R^2}
= \Omega^2 {3 r_0^2 - 4 R^2 \over 2 r_0^2 - R^2}.
\label{eq14}
\end{equation}
The orbits are unstable for 
\begin{equation}
z_0 < \sqrt{3} R,
\label{eq15}
\end{equation}
or equivalently, for
\begin{equation}
r_0 < {2 \sqrt{3} \over 3} R = 1.1547 R \equiv r_{\rm crit}.
\label{eq16}
\end{equation}
As $r_0$ approaches $R$, the instability growth time approaches
an orbital period.

Another example is the Gaussian surface of revolution,
\begin{equation}
r^2 = R^2 e^{z^2},
\label{eq17}
\end{equation}
which has a minimum radius $R$, and a critical radius $r_{\rm crit}  = R
\sqrt[4]{e} = 1.28 R$.

Our final example is the surface
\begin{equation}
r = - {k \over z \sqrt{1 - z^2}}\ , \qquad (-1 < z < 0),
\label{eq18}
\end{equation}
which has a minimum radius of $R = 2k$, approaches the surface $r = -k/z$ at 
large $r$ (small $z$), and has a critical radius of $r_{\rm crit} = 
6k/\sqrt{5} = 1.34 R$.

These examples arise in a $2 + 1$ geometry with curved space but flat time.
As such, they are not fully analagous to black holes in $3 + 1$ geometry with
both curved space and curved time.  Still, they provide a glimpse as to
how a particle in curved spacetime can undergo considerably more complex 
motion than in flat spacetime.

\section{Acknowledgement}

The author wishes to thank Ori Ganor and Vipul Periwal for
discussions of this problem.

\end{document}